\documentclass{elsart3p}
\usepackage{psfig,graphics,graphicx}
\usepackage{color}
\usepackage{amssymb}
\usepackage{bookmath}
\usepackage{amssymb}

\begin{document}

\begin{frontmatter}



\title{Anisotropy of heat capacity in Pauli limited unconventional superconductors\thanksref{BR}}
\thanks[BR]{Supported by the Board of Regents of Louisiana}


\author{A. Vorontsov},
\author{I. Vekhter}

\address{Department of Physics and Astronomy,
             Louisiana State University, Baton Rouge, Louisiana, 70803, USA}

\begin{abstract}
Diamagnetically coupled magnetic field can be used as a probe of nodal positions 
in unconventional superconductors. The heat capacity depends on the 
angle $\phi_0$ between the magnetic field and the nodal directions.
We show that the anisotropy $C(\phi_0)$ persists even in systems with strong 
paramagnetic coupling to the electrons's spins.
\end{abstract}

\end{frontmatter}

Anisotropy of thermodynamic properties of a superconductor in 
rotated magnetic field
is often used to determine the location of 
the nodes in the superconducting gap at the Fermi surface. 
Magnetic field serves as a directional probe since 
the circulating screening supercurrents around vortices 
excite quasiparticles differently, 
depending on the orientation of the field with respect to the 
nodes. While initial work 
suggested that the density of states and the specific heat have 
a minimum
when the field is applied along a node, we recently argued 
that the situation is more complex, and minima and maxima
interchange in the $T$-$H$ phase diagram  
\cite{IVekhter:1999R,AVorontsov:2006}.

Several superconducting materials, including heavy-fermion 
CeCoIn$_5$, have strong suppression of superconductivity 
due to paramagnetic coupling of electron spins to the field. 
This effect is insensitive to the field orientation and competes 
with the orbital coupling. 
We investigate how the anisotropy of the heat capacity is 
affected by this competition.

We follow the quasiclassical method \cite{AVorontsov:2006,AHoughton:1998} 
and 
include interaction of electron spins with the field via Zeeman term.
The  $4\times4$(spin-up-down-particle-hole) 
Green's function $\whg(\vR, \hat{\vp}; \vare)$
satisfies Eilenberger equation\cite{qc} 
\be
[\vare \widehat{\tau}_3 - \widehat{v}_{orb} - \widehat{v}_Z-\whDelta - \whs_{imp}, 
\whg] + i\vv_f\cdot \grad \; \whg = 0 \,. 
\label{eq:eilZ}
\ee
Here $\whDelta(\vR, \hvp)$ is the mean-field order parameter, $\whs_{imp}(\vR; \vare)$ is 
the impurity self-energy. 
The orbital coupling of magnetic field is via the vector potential $\vA(\vR)$, 
while the Zeeman term couples $\vB$ to the magnetic moments of electrons $\mu = (g/2) \mu_B$, 
$$
\widehat{v}_{orb} = - {e\over c} \vv_f \vA \; \widehat{\tau}_3 \qquad 
\widehat{v}_Z = \left( \begin{array}{cc} 
\mu\vsigma\cdot\vB & 0 \\ 0 & \mu\vsigma^*\cdot\vB \end{array} \right) \,.
$$
In the absence of spin-orbit interaction we choose the direction of 
the field to be the spin quantization axis. Then 
all matrices 
split into independent blocks, for spin up and down. 
The equations for the two spin directions are independent, and differ 
only by a spin-dependent energy shift. For example, the equation for 
the off-diagonal components of $\whg(\vR, \hat{\vp}; \vare)$ is
\bea
\left[-2i(\tilde{\vare}\mp\mu B) + \vv_f(\hat{\vp}) \left(
\gradR - i \frac{2e}{\hbar c} \vA(\vR) \right) \right] \, \times
\nonumber \\
f^R_{\uparrow,\downarrow}(\hat{\vp}, \vR; \vare)
 = 2 \tilde{\Delta} \, i g^R_{\uparrow,\downarrow}(\hat{\vp}, \vR; \vare).
\eea
Here the energy $\tilde{\vare}=\vare-\Sigma_{\uparrow,\downarrow}$ and 
order parameter $\tilde{\Delta} = \Delta + \Delta_{imp, \uparrow,\downarrow}$ 
are renormalized by impurity self-energies.
In analogy to Ref.\cite{AVorontsov:2006}, we take the $d_{x^2-y^2}$ gap 
$\Delta = \Delta(\vR) \sqrt{2} \cos2\phi$, 
with spatial structure given by Abrikosov vortex lattice, and do not 
consider a possible additional modulation due to 
Fulde-Ferrell-Larkin-Ovchinnikov state.
We solve these equations for Green's function using 
Brandt-Pesch-Tewordt approximation.\cite{BPT}
To these we add self-consistency equations on $\whs_{imp}$ and $\Delta$. 
The latter is the only place where two spins add, 
$\Delta(\vR,\hvp) = \int {d\vare\over 4\pi i} \tanh{\vare\over2T} 
\int \d\hvp' V(\hvp,\hvp') \onehalf (f_\uparrow + f_\downarrow)$.

The specific heat is calculated from the density of states for two spins 
$N_{\uparrow,\downarrow}$ using,
$$
C(T,\phi_0) = \int\limits^{+\infty}_{-\infty}
\; d\vare \; \frac{\vare^2 \; (N_\uparrow(T, \phi_0; \vare) + N_\downarrow(T, \phi_0; \vare) )}
{4 T^2 \cosh^2(\vare/2T)} \,, \nonumber
$$
which is valid at low temperature, $T \ll T_c$. 
The calculations are done for a quasi-cylindrical Fermi surface, 
$p_f^2 = p_x^2 + p_y^2 - r^2 p_f^2 \cos(2s p_z/r^2 p_f)$, with 
parameters $r=s=0.5$ that ensure the 3D nature of the vortices.\cite{AVorontsov:2006} 
Parameter $Z=\mu B_0/2\pi T_c$, where $B_0 = (ch/2e)/2\pi \xi_0^2$,
and the in-plane coherence length $\xi_0 = \hbar v_f/2\pi T_c$,
characterises the 
strength of the Zeeman term.

\begin{figure}[t]
\centerline{\includegraphics[height=5cm]{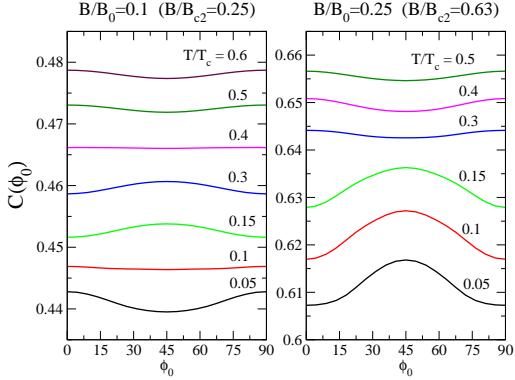}}
    \caption{\label{fig:hcap} The heat capacity anisotropy for $Z=0.4$ 
at different temperatures.  Anisotropy curves correspond to points indicated 
in the right panel of Fig.~\ref{fig:PD}. 
(The upper curves are shifted down to be on the same scale.) 
}
\end{figure}
Figure \ref{fig:hcap} shows the anisotropy of the specific heat 
as a function of $T$ and $B$.
For $\vB$ in the nodal direction, $\phi_0=45^o$, $C(\phi_0)$
has minima or maxima depending on the temperature and field range.
This is similar to the case of purely orbital coupling, $Z=0$, and
is due to the interplay between excitation and scattering of
quasiparticles by magnetic field, which is affected by 
the quasiparticles energy ($\sim T$), and magnitude and the 
orientation of the field.\cite{AVorontsov:2006}
\begin{figure}[t]
\centerline{\includegraphics[height=5cm]{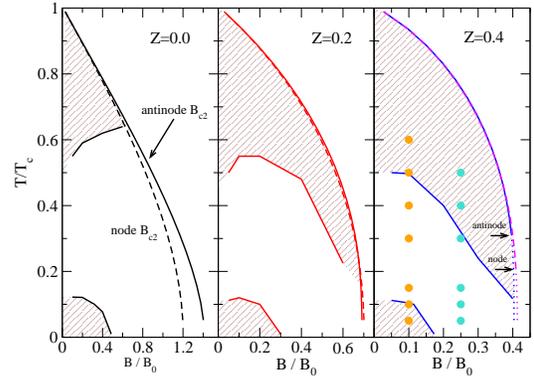}}
    \caption{\label{fig:PD} Phase diagram of the $C(T,\phi_0)$
anisotropy  
for different strength of the paramagnetic term. In the
shaded areas $C(\phi_0)$ has a minimum when the field is along a
nodal direction. 
The arrows in the right panel indicate onset of the first order transition 
occurring for $Z^* \gtrsim 0.35$.
}
\end{figure}

Fig.~\ref{fig:PD}  shows
regions of minima and maxima at $\phi_0=45^o$ in the $T$-$B$
plane, and their evolution with 
increasing Zeeman term $Z$.
Of the two areas where $C(\phi_0)$ has a minimum at 45$^\circ$,
the low-$T$, low-$B$ regime is only weakly affected by variation in $Z$.
In contrast, the region at high $T$ expands
as we increase $Z$, at the expense of the
area where a maximum at $\phi_0=45^o$ is observed. Note that the in-plane anisotropy of $B_{c2}$ (present for $Z=0$) quickly disappears as 
we turn on the strength of the Pauli coupling. 

The anisotropy of $C(\phi_0)$ is usually measured below $T_c/2$
\cite{TPark:2003,HAoki:2004}, 
where the effect of the Zeeman coupling is weak. While increased $Z$ expands the high-$T$ ``minimum'' region, the amplitude of the anisotropy there is probably still below the experimental
resolution. We conclude that the Zeeman coupling has weak effect on the observed anisotropy.




%
%
%
%
%

\end{document}